\documentclass[a4paper,12pt]{article}
\usepackage{epsfig}
\date{\today}

\newcommand{\bmat}{\left(\begin{array}}
\newcommand{\emat}{\end{array}\right)}
\newcommand{\be}{\begin{equation}}
\newcommand{\ee}{\end{equation}}
\newcommand{\ba}{\begin{eqnarray}}
\newcommand{\ea}{\end{eqnarray}}

\def\lsim{\raise0.3ex\hbox{$\;<$\kern-0.75em\raise-1.1ex\hbox{$\sim\;$}}}
\def\gsim{\raise0.3ex\hbox{$\;>$\kern-0.75em\raise-1.1ex\hbox{$\sim\;$}}}

\def\be{\beta}

\def\lsim{\raise0.3ex\hbox{$\;<$\kern-0.75em\raise-1.1ex\hbox{$\sim\;$}}}
\def\gsim{\raise0.3ex\hbox{$\;>$\kern-0.75em\raise-1.1ex\hbox{$\sim\;$}}}

\begin{document}

\renewcommand{\thefootnote}{\fnsymbol{footnote}}

%\begin{titlepage}
%\pagestyle{empty}
\rightline{KUNS-1826}
\vskip 1cm
\begin{center}
{\bf \large{  Heterotic Yukawa couplings and continuous Wilson lines
\\[10mm]}}
{ Tatsuo Kobayashi $^{1}$ and Oleg Lebedev $^{2}$\\[6mm]}
\small{$^1$Department of Physics, Kyoto University, Kyoto 606-8502,~~Japan\\[2mm]}
\small{$^2$Centre for Theoretical Physics, University of Sussex, Brighton BN1
9QJ,~~UK\\[2mm]}
\end{center}

\hrule
\vskip 0.3cm
\begin{minipage}[h]{14.0cm}
\begin{center}
\small{\bf Abstract}\\[3mm]
\end{center}

Realistic heterotic string models require the presence of
the Wilson lines.
We study the effect of continuous Wilson lines on the heterotic Yukawa couplings for abelian orbifold compactifications  and find that  
the presence of continuous Wilson lines 
affects the magnitude of the twisted Yukawa couplings
resulting in their stronger hierarchy.

\end{minipage}
\vskip 0.7cm
\hrule
\vskip 1cm

{\bf 1. Introduction.}

\vskip 0.5cm

One of the ultimate tests of string theory is the prediction
of the quark and lepton masses. This necessitates the
knowledge of the Yukawa couplings in string models.
In the context of the heterotic string \cite{Gross:1984dd}, which provides many
attractive models, the desired Yukawa couplings have been
explicitly calculated for the orbifold compactifications.
The Yukawa couplings of the twisted sectors are generated
by the world sheet instantons and can be computed using 
conformal field theory techniques.
The original calculation was performed by Dixon $et$. $al$. \cite{Dixon:1986qv},
and Hamidi and Vafa \cite{Hamidi:1986vh}, and later refined in \cite{Burwick:1990tu},\cite{erler}.
The Yukawa couplings were found to be exponentially suppressed
by the distance between the orbifold fixed points where
the twisted fields reside. This offered a natural explanation of 
 a 
hierarchical pattern of the quark and lepton mass matrices
(see \cite{Ibanez:1986ka}   for  related studies).
The effect of the antisymmetric background field $B_{ij}$ 
was included in \cite{erler}.
This had important implications for the studies of CP violation
in the context of string theory. In particular, it was realized
that the antisymmetric background field may lead to complex
Yukawa couplings \cite{Bailin:1998xx} and eventually to CP violation in the
Cabibbo--Kobayashi--Maskawa matrix 
\cite{Lebedev:2001qg}\footnote{For analogous results in Type I
models, see \cite{Aldazabal:2000sa}.}.

In addition to  $B_{ij}$, there may exist other
backgrounds such as the Wilson lines  \cite{Narain:am}. 
In fact, their presence is necessary to produce reasonable
phenomenology, e.g. the Standard Model gauge group (or close to such) and 
a small number of generations \cite{Ibanez:1986tp},\cite{Ibanez:1987xa}.
The missing link so far has been the effect of the Wilson lines on
the Yukawa couplings. 
Wilson lines can be continuous or discrete depending on the embedding
of the space group into the gauge degrees of freedom. 
Realistic models involve discrete or, possibly, both types 
of the Wilson lines. 
In this Letter, we study continuous Wilson lines and defer
the analysis of discrete Wilson lines 
until  our subsequent publication.

The presence of background fields generally
breaks the CP symmetry (as defined in 
Refs.\cite{Dine:1992ya,Kobayashi:1994ks}  
in the context of string theory).
So, an important question to address is whether or not this CP violation
shows up in the Yukawa couplings.
In the case of the intersecting brane models, this issue 
has been considered in Ref.\cite{Cremades:2002va}
   with the result that
the Wilson lines induce CP violating Yukawa couplings.
In heterotic models, the problem is more complicated. In this work,
we compute the Wilson line contribution to the Yukawa couplings
for orbifold compactifications admitting the decomposition
$T^2 \oplus T^2 \oplus T^2$.

We start with reviewing the technique of computation of the Yukawa couplings.

\vskip 1cm

{\bf 2. $T$--moduli dependence of the Yukawa couplings.}

\vskip 0.5cm

A $Z_N$ orbifold is  defined as  a torus modded by a twist $\theta$
which acts as a discrete rotation of order $N$ (for a review, see
\cite{Bailin:nk}).
Strings closed on the original torus correspond to the untwisted
sector, while those closed only on the orbifold form twisted sectors.
Twisted states have no momentum in the compactified directions
and have their center of mass at one of the fixed points $f$
defined by
\begin{equation}
\theta f =f + \Lambda ~,
\end{equation}
where $\Lambda$ is a torus lattice vector.
Below we consider Yukawa couplings of the twisted states. 
They depend on various moduli such as compactification radii,
Wilson moduli, etc. In contrast,  Yukawa couplings
of the untwisted states are constant and therefore irrelevant
for our purposes.

A twisted trilinear coupling of two fermions and a boson 
corresponds to a string amplitude for massless states
\begin{equation}
\langle  V^B_{\alpha, f_\alpha}  V^F_{\beta, f_\beta} 
V^F_{\gamma, f_\gamma}  \rangle \;,
\label{vertex}
\end{equation}
where $\alpha,\beta,\gamma$ are associated with twists
$k/N,l/N,-(k+l)/N$ of a $Z_N$ orbifold and $f_{\alpha,\beta,\gamma}$
label orbifold fixed points. The twisted vertex operators are defined
through the twist fields $\sigma_{\alpha, f_\alpha}$:
\begin{equation}
V_{\rm tw}(z,\bar z)= V_{\rm untw}(z,\bar z)~\sigma_{\alpha, f_\alpha}(z,\bar z)
\end{equation}
and $\sigma_{\alpha, f_\alpha}$ creates a twisted ground state from an
untwisted  vacuum \footnote{For higher twisted sectors, the twist fields 
$\sigma_{\alpha, f_\alpha}$ themselves, in general, 
do not correspond to eigenstates of $\theta$.
One has to take their linear combinations to obtain 
physical states \cite{Kobayashi:1990mc,Kobayashi:1991rp}.},  
$\sigma_{\alpha, f_\alpha}(z,\bar z) \vert 0 \rangle =
\vert 0_{\alpha, v} \rangle $ with $v=(1-\alpha)(f_\alpha + \Lambda)$ .

The vertex operator for the emission of an untwisted massless state
with momentum $P$ is given by
\begin{equation}
V(z, \bar z)=e^{-q \Phi(z)} e^{i \alpha \cdot H} e^{i(P_R \cdot X_R-
P_L \cdot X_L)} \;,
\end{equation}
where $q \Phi(z)$ is the superghost contribution,
$\alpha$ is an $SO(10)$ weight vector, and $H(z)$ 
represents bosonized NSR fermions. Since twisted strings have no momentum
in the orbifold directions, the path integral in Eq.(\ref{vertex})
factorizes:
\begin{equation}
\langle  V^B_{\alpha, f_\alpha}  V^F_{\beta, f_\beta} 
V^F_{\gamma, f_\gamma}  \rangle =
\langle  V^B  V^F  V^F  \rangle \langle  
 \sigma_{\alpha, f_\alpha}   \sigma_{\beta, f_\beta} \sigma_{\gamma, f_\gamma} \rangle \;.
\end{equation}
In the second factor the path integral is to be performed over six
``internal'' string fields. The twisted Yukawa coupling is then given 
by a low--energy limit of this  expression integrated over all
possible vertex locations $z_{1,2,3}$. We note that the first factor is 
needed to make the result $SL(2,C)$ invariant, while the second factor
contains the orbifold information.
Furthermore, only the $classical$ part of $X^i$ provides the moduli 
dependence relevant for our purposes.

Thus, the  Yukawa couplings among the twisted sectors $\alpha,\beta,\gamma$ 
can be expressed as 
\begin{equation}
Y_{\alpha \beta \gamma}= {\rm const}~ \sum_{X_{\rm cl}} e^{-S_{\rm cl}} \;,
\end{equation}
where $X_{\rm cl}$ are solutions to the string equations of motion in the
presence of the twist fields 
$\sigma_{\alpha, f_\alpha}, \sigma_{\beta, f_\beta}, \sigma_{\gamma, f_\gamma}$ located
at points $z_{1,2,3}$ of the world sheet.
With the normalization of  Ref.\cite{Bailin:nk}, the classical action is\footnote{Note the difference in the definitions of $Z^i$,
see Eqs.(3.47) and (1.48).}
\begin{equation}
S_{\rm cl}=
{1\over 2\pi} \int d\tau d\sigma ~\partial^\alpha X^{\rm i} 
 \partial_\alpha  X^{\rm i}
={1\over 2\pi} \int d^2 z (\partial Z^i \bar \partial \bar Z^i+
\bar\partial Z^i \partial \bar Z^i ) 
\end{equation}
with $Z^i=X^i+iX^{i+1}$, $\bar Z^i=X^i-i X^{i+1}$   and $z=e^{-2(\tau +i\sigma)}$. Henceforth, roman indices label the compactified directions,
i=1,..,6, while italic indices label the compactified planes,
$i$=1,3,5.
The resulting equations of motion are
\begin{equation}
\partial^2 Z^i/ \partial z \partial \bar z =0 \;,
\end{equation}
such that $Z^i$ can be split into a holomorphic and an antiholomorphic pieces,
$Z^i=Z^i_R(z) + Z^i_L(\bar z)$.
The holomorphic (right--moving) piece $Z_R$ belonging to the
twisted sector $k/N$ can be expanded as
\begin{equation}
Z_R=z_R + {i\over 2} \sum_{n=1}^\infty {\beta_{n-k/N} \over n-k/N}
~z^{-(n-k/N)}-
{i\over 2} \sum_{n=0}^\infty {\gamma_{n+k/N}^\dagger \over n+k/N}
~z^{n+k/N} \;,
\end{equation}
where $z_R$ is a fixed point, and 
$\beta$ and $\gamma^\dagger$ are  the annihilation and creation operators.
For $z \rightarrow 0$,
\begin{equation}
\partial_z Z \vert 0_{k/N} \rangle \rightarrow -{i\over 2} z^{k/N-1}
 \gamma_{k/N}^\dagger  \vert 0_{k/N} \rangle\;,
\end{equation}
so we have the following OPE
\begin{equation}
\partial_z Z \sigma_{k/N}(0,0) ~\sim~ z^{-(1-k/N)} \tau_{k/N}(0,0)\;,
\end{equation}
where the operator $\tau_{k/N}$ creates an excited twisted state from
an untwisted vacuum.
Therefore, the classical solutions with the correct behaviour at the
twist insertion points $z_{1,2,3}$ are of the form
\begin{eqnarray}
&&\partial Z =c  ( z- z_1)^{-(1-k/N)} 
( z- z_2)^{-(1-l/N)} ( z- z_3)^{-k/N-l/N}\;,\nonumber\\
&&\bar \partial \bar Z =\bar c (\bar z-\bar z_1)^{-(1-k/N)} 
(\bar z-\bar z_2)^{-(1-l/N)} (\bar z-\bar z_3)^{-k/N-l/N}\;,\nonumber\\
&&\bar \partial Z =d (\bar z-\bar z_1)^{-k/N} 
(\bar z-\bar z_2)^{-l/N} (\bar z-\bar z_3)^{-(1-k/N-l/N)} \;,\nonumber\\
&& \partial \bar Z =\bar d ( z- z_1)^{-k/N} 
( z- z_2)^{-l/N} ( z- z_3)^{-(1-k/N-l/N)} 
\end{eqnarray}
for each complex plane. The constants $c,d$ are to be determined by the
boundary conditions for $Z^i$. In particular, encircling the twist
insertion points an appropriate number of times produces a winding vector.
Therefore, we have the following $monodromy$ conditions
\begin{eqnarray}
\Delta Z^i= \int_{\cal C} dz ~\partial Z^i + \int_{\cal C} d\bar z ~
\bar \partial Z^i= v^i \;, \nonumber\\
\Delta \bar Z^i= \int_{\cal C} dz ~\partial \bar Z^i + \int_{\cal C} d\bar z ~
\bar \partial \bar Z^i= \bar v^i \;,
\end{eqnarray}
with the complex lattice vector $v^i$ defined by $v^i=v^{(i)} +i v^{(i+1)}$
and $\bar v^i=v^{(i)} -i v^{(i+1)}$.
Here the contour ${\cal C}$ is chosen such that $Z^i$ gets shifted but
not rotated upon going around ${\cal C}$.
These equations allow to solve for $c,d$ in terms of the winding
vectors $v^i$.

In the case of a three point correlator, $z_{1,2,3}$ can be 
transformed into  $0$, $1$, and $\infty$ by  an appropriate $SL(2,C)$
transformation. 
By encircling point $0$ $l$ times clockwise and point 1 $k$ times
anticlockwise, $Z^i$ is shifted by
\begin{equation}
v^i=(1-\alpha^l)(f_\alpha - f_\beta +\Lambda)\;.
\label{v^i}
\end{equation}
This follows from the space group multiplication rule for
$(\alpha, { l}_{\alpha})^l (\beta, { l}_{\beta})^{-k}$
with ${ l}_{\alpha}=(1-\alpha)(f_\alpha +\Lambda)$
and ${ l}_{\beta}=(1-\beta)(f_\beta +\Lambda)$.
Encircling all three points produces a contractable  cycle,
$(\alpha,{ l}_{\alpha} )(\beta,{ l}_{\beta} )
(\gamma,{ l}_{\gamma} )=(1,0) $,
 hence we obtain the point group selection rule $\alpha\beta\gamma=1$ and 
the space group selection rule\footnote{See \cite{Kobayashi:1991rp} 
for discussion.}
\begin{equation}
(1-\alpha) f_\alpha + (1-\beta) f_\beta +(1-\gamma) f_\gamma=0
\end{equation}
up to a lattice vector.

Typically there is only one independent nontrivial contour ${\cal C}$ 
in the monodromy condition \cite{Dixon:1986qv}
(yet, there may be
subleties discussed in Ref.\cite{erler}), while other contours can be obtained by
adding  ${\cal C}$ and trivial contours which encircle one point
$n$ times or all three points.
Correspondingly, there is  one equation
for two variables $c$ and $d$. The case with non-zero $d$ does not contribute
to the Yukawa couplings due to a divergence of the action ($S_{\rm cl}
\rightarrow +\infty$). Thus, we set $d=0$ and find (for each complex plane)
\begin{equation}
c={i(-z_\infty)^{(k+l)/N} \over 2} {\Gamma ((k+l)/N)~v  \over
\sin (k l \pi/N) \Gamma (k/N) \Gamma (l/N)}\;.
\end{equation} 

The Yukawa couplings are determined by the holomophic instantons, i.e.
classical solutions with holomorphic $Z^i$ and antiholomorphic $\bar Z^i$
(yet, this does not imply that $X^i$ is purely right- or left--moving). 
Performing the $d^2z$ integral using the method of Kawai et. al. 
\cite{Kawai:1985xq},
we obtain \cite{Bailin:nk}
\begin{equation}
S_{\rm cl}(v_i)= {\vert v_i\vert^2  \over 4\pi \sin^2(k_i l_i \pi/N)}
{\vert \sin(k_i \pi/N)\vert \vert \sin(l_i\pi/N)\vert \over
\vert \sin((k_i+l_i)\pi/N)\vert} \;.
\end{equation}
The sum over $X_{\rm cl}$ in the correlation function reduces to the 
sum over $v_i$ which is parametrized by a lattice vector $\Lambda$
through (\ref{v^i}). The $T$--moduli dependence is contained in
$\vert v_i\vert^2$ which is proportional to the square
of the relevant compactification radius, and Re$T_i \propto R_i^2$.
\vskip 1cm

{\bf 3. Effect of continuous Wilson lines.}

\vskip 0.5cm 

A continuous Wilson line is realized through the correspondence
between the space group and rotations and shifts of the gauge
$E_8 \times E_8$ lattice \cite{Ibanez:1987xa}:
\begin{equation}
(\theta, l) ~  \longrightarrow ~ (\Theta,a)  ~~,
\end{equation}
where $\theta$ is a point group element, $\Theta$ is a rotation
of the $E_8 \times E_8$ lattice, and $l$ and $a$ are related by
\begin{equation}
l^{\rm i}=\sum_\alpha n_\alpha {e_\alpha^{\rm i}} \;\;,\;\;
a^{\rm I}=\sum_\alpha n_\alpha A_\alpha^{\rm I} \;.
\end{equation}
Here $n_\alpha$ are some integers, $e_\alpha^{\rm i}$ are the torus basis vectors,
and $A_\alpha^{\rm I}$ are the Wilson lines.
In contrast to the discrete Wilson lines, 
$a^{\rm I}$ are unconstrained as long as they are rotated by $\Theta$
since $(\Theta, a)^N=(1,0)$ automatically.
Note that if we identify $\stackrel{\rightarrow}{e_1} $ with 
$\stackrel{\rightarrow}{a_1}$, the group multiplication rule
$(\theta_2,l_2) (\theta_1, l_1)=(\theta_2 \theta_1, l_2 +\theta_2 l_1)$
requires us to identify $\theta \stackrel{\rightarrow}{e_1}$ 
with $\Theta \stackrel{\rightarrow}{a_1}$.
In particular, if $\theta \stackrel{\rightarrow}{e_1}=
\stackrel{\rightarrow}{e_2}$, then $\Theta \stackrel{\rightarrow}{a_1}=
\stackrel{\rightarrow}{a_2}$.

The presence of a Wilson line $A_{\rm iI}$ gives an additional contribution to the action \cite{Narain:am}
\begin{equation}
\Delta S_{\rm cl}= {A_{\rm iI}\over 2\pi} \int d\tau d\sigma ~
\epsilon^{\alpha\beta} \partial_\alpha X^{\rm i} \partial_\beta X^{\rm I}\;,
\label{action}
\end{equation}
with I=1,..,16 labelling the left--moving gauge space coordinates
$X^{\rm I}$ .
Summing up the contributions from $A_{iI}~,~A_{i+1,I}~,~A_{i,I+1}~,~A_{i+1,I+1}$
and using  the complex coordinate $z=e^{-2(\tau +i\sigma)}$, 
we write the $Euclidean$ action as
\begin{equation}
\Delta S_{\rm cl}= {1\over 2\pi }\int d^2z \biggl[ 
 {\cal A}_{iI} (   \partial Z^i \bar \partial   Z^I     -
\bar\partial Z^i \partial   Z^I) +
{\cal A}_{iI}' (   \partial Z^i \bar \partial   \bar Z^I        -
\bar\partial Z^i \partial   \bar Z^I) \biggr] - {\rm h.c.} \;,
\end{equation}
where h.c. replaces a quantity with the corresponding barred one
and conjugates ${\cal A},{\cal A}'$, and
we have defined 
\begin{eqnarray}
&& Z^I=X^I+iX^{I+1} \;\;, I=1,3,..,15 \;, \nonumber\\
&&  {\cal A}_{iI}= {1\over 4}(A_{iI}-A_{i+1,I+1}-iA_{i+1,I}-iA_{i,I+1}) 
                                \;,\nonumber\\
&&  {\cal A}_{iI}'= {1\over 4}(A_{iI}+A_{i+1,I+1}-iA_{i+1,I}+iA_{i,I+1}) \;.
\end{eqnarray}
The expression for the action can be simplified using the fact that
$X^{\rm I}$ is a left mover, $\partial \bar Z^I \sim \partial Z^I \sim (\partial_\tau - \partial_\sigma)X^{\rm I}=0$.

The addition of the Wilson line does not affect the equations
of motion for $X^{\rm i}$ since it generates only a boundary term contribution.
Nor does it affect the monodromy conditions because the boundary conditions
for $X^{\rm i}$ remain intact. As a result, the classical solutions for $Z^i$
are the same as in the case without Wilson lines.

We also note that the pure gauge space contribution  to the action
\begin{equation}
\Delta S_{\rm gauge}={1\over 2\pi} \int d^2 z ~(\partial Z^I \bar \partial \bar Z^I+
\bar\partial Z^I \partial \bar Z^I )\;. 
\end{equation}
vanishes when the left--mover constraint is applied.
Thus, the only effect of the Wilson lines on the Yukawa couplings comes
from the additional term (\ref{action}) in the world sheet action. 

The Wilson lines are subject to certain constraints \cite{Mohaupt:1993fb}. In particular,
the invariance of the action (\ref{action})  under twisting requires
(in orthogonal coordinates)
\begin{equation}
\theta^T A~ \Theta= A  \;.
\end{equation}
A common choice (the ``standard embedding'') is $\Theta=\theta$
which acts on a 6D sublattice of the $E_8 \times E_8$ lattice, while
leaving the other components intact.
For a twist of any order, this implies\footnote{In the convention of  Ref.\cite{Cvetic:1995dt}, 
our Wilson matrix $A$ corresponds to
$(U^T C A U)^T$ up to an overall normalization,  
where C is the $E_8 \times E_8$ Cartan metric and $U^T CU=$diag(2,2,..,2). } 
\begin{equation}
A= \left( \matrix{a & -b \cr
                  b & a }
\right)
\end{equation}
for each  of the three planes.
Here $a,b$ are real continuous moduli. As a result, ${\cal A}_{iI}=0$ and 
\begin{equation}
\Delta S_{\rm cl}= {{\cal A}_{iI}' \over 2\pi }\int d^2z   
~ \partial Z^i \bar \partial   \bar Z^I  \;,
\end{equation}
where we have applied the left--mover constraint
and used $\bar \partial Z^i=  \partial \bar Z^i =0$.

In the case of the standard embedding, $\bar\partial \bar Z^i$ and 
$\bar\partial\bar Z^I$ have the same 
structure. In particular, they have the same singular behaviour
at the twist operator insertion points:
\begin{equation}
\bar\partial\bar Z^I= \bar c' (\bar z-\bar z_1)^{-(1-k/N)} 
(\bar z-\bar z_2)^{-(1-l/N)} (\bar z-\bar z_3)^{-k/N-l/N}\;.
\end{equation}
The constant $c'$ is determined by the monodromy condition for $Z^I$:
\begin{eqnarray}
\int_{\mathcal C} d\bar z~ \bar \partial \bar Z^I = \bar u \;,
\end{eqnarray}
with the result
\begin{equation}
c'= c ~{u\over v}\;.
\end{equation}
Here $u$ is the gauge space representation of the space group element
$v$ which appears in the monodromy condition for $Z^i$.
For example, 
in the basis $\{\stackrel{\rightarrow}{e_1},
\stackrel{\rightarrow}{e_2}=\theta \stackrel{\rightarrow}{e_1} \}$   this correspondence is given  (in each plane) 
by\footnote{Here we have set the compactification radii to one.
In general, these expressions are valid for $v/R$ and $u/R$.} 
\begin{eqnarray} 
&& \stackrel{\rightarrow}{v}= n_1 \stackrel{\rightarrow}{e_1}
+ n_2 ~\theta \stackrel{\rightarrow}{e_1} ~\longrightarrow ~ v=
n_1 + n_2 ~e^{2\pi i/N}  \;,\nonumber\\
&& \stackrel{\rightarrow}{u}=  n_1 \stackrel{\rightarrow}{a_1}
+ n_2 ~\theta \stackrel{\rightarrow}{a_1} ~\longrightarrow ~ u= 
n_1  (a-ib) +  n_2~ e^{2\pi i/N} (a-ib) \;.
\end{eqnarray}
Here $\stackrel{\rightarrow}{a_1}=(a,-b)$.

The remaining integral $\int d^2z~\vert \partial Z^i  \vert^2$ is identical to the one appearing
in the calculation of the Yukawa couplings without the Wilson lines.
The only difference arises due to the factor
\begin{equation}
{\cal A}_{11}' v \bar u ={1 \over 2} \Bigl(a^2 +b^2\Bigr) 
\vert v \vert^2  
\end{equation}
for each plane. 
Here $v$ is the space group element from the monodromy condition.
As a result,
\begin{equation}
S_{\rm cl}(v)= {\vert v\vert^2  \over 4\pi \sin^2(kl\pi/N)}
{\vert \sin(k\pi/N)\vert \vert \sin(l\pi/N)\vert \over
\vert \sin((k+l)\pi/N)\vert}  \left(1+ {a^2 +b^2 \over 2}\right)
\end{equation}
for each plane.

Turning on the antisymmetric background field $B_{ij}$
\cite{Bailin:nk}, we obtain the 
following Yukawa couplings
\begin{eqnarray}
&&Y_{\alpha\beta\gamma} = {\rm const } \times \label{result} \\ 
&&\sum_v \exp \biggl[-\sum_{i=1,3,5}
{\vert v_i \vert^2  \over 4\pi \sin^2(k_i l_i \pi/N)}
{\vert \sin(k_i \pi/N)\vert \vert \sin(l_i\pi/N)\vert \over
\vert \sin((k_i +l_i)\pi/N)\vert}  \left(1-iB_{i,i+1}  + {1\over 2} a_i \bar a_i\right) \biggr] \;, \nonumber 
\end{eqnarray}
 where we have defined $a_i=A_{i,i}+i A_{i,i+1}$.
Clearly, this expression is invariant under rotations of the Wilson lines.
We see that, for $B_{ij}=0$, the presence of the Wilson lines   amounts essentially
to a rescaling of $v_i$ or the compactification radii. This creates a $stronger$
hierarchy of the Yukawa couplings of the fields placed at
different fixed points of the orbifold.

These results can be generalized to a class of non--standard embeddings
such that a $Z_N$ twist is associated with rotations of more than one
planes in the gauge space, $(\theta,l) \rightarrow (\Theta_{(1)},a_{(1)}),
(\Theta_{(2)},a_{(2)}),...$
In this case $A$ can be split into the $2\times 2$ blocks
\begin{equation}
A_{(1)}= \left( \matrix{a_{(1)} & -b_{(1)} \cr
                  b_{(1)} & a_{(1)} }
\right)\;\;,\;\;
A_{(2)}=\left( \matrix{a_{(2)} & -b_{(2)} \cr
                  b_{(2)} & a_{(2)} }
\right)\;\;,...
\end{equation}
This results in the substitution
\begin{equation}
{1\over 2} a_i \bar a_i ~ \longrightarrow ~ {1\over 2} \sum_{k} [a_{(k)}]_{_i} 
 [\overline a_{(k)}]_{_i}
\end{equation}
in Eq.(\ref{result}).

In terms of the T--moduli \cite{Bailin:nk},
\begin{equation}
 \vert v_i \vert^2 \left(1-iB_{i,i+1}  + {1\over 2} a_i \bar a_i \right) ~ \propto ~
T_i +\kappa (T_i +\bar T_i) a_i \bar a_i \;,
\end{equation}
where $\kappa$ is a constant depending on the orbifold.
The effect of the axionic shift $T_i \rightarrow T_i + i$  on the Yukawa couplings is 
therefore the same as in the case without the Wilson lines.
For phenomenological applications it implies that
the Cabibbo-Kobayashi-Maskawa (CKM) phase vanishes for 
Im$T_i=\pm 1/2$ and at the fixed points of the modular group,
in particular \cite{Lebedev:2001qg} (see also \cite{Dent:2001cc}). This follows from the fact that the 
Jarlskog invariant $J$=ImDet$[Y_u Y_u^\dagger,Y_d Y_d^\dagger ]$
satisfies \cite{Lebedev:2001qg}
\begin{equation}
J(T_i)=J(T_i +i)=-J(T_i)=0
\end{equation}
for Im$T_i=\pm 1/2$. The Jarlskog invariant serves as an indicator of 
CP violation in the Standard Model such that its vanishing
implies a zero CKM phase. This means that if the $T$--moduli
are the source of the observed CP violation, they cannot be
stabilized at Im$T_i=\pm 1/2$. This requirement puts rather
severe constraints on heterotic string models \cite{Khalil:2001dr}.

A curious feature of this result is that the Wilson line
background in heterotic models does not lead to CP violating 
Yukawa couplings. This is in contrast to the open string case 
\cite{Cremades:2002va},
in which the Wilson lines provide phase factors in the Yukawa
matrices and may in principle  be a source of CP violation.
The difference arises due to the  embedding of the space group
into the gauge group via the Wilson line, which is not implemented
in the open string case\footnote{Yet, the effect of $B_{ij}$
is CP violating in both cases.}. 
Even though the Wilson line background may break CP,
this sort of CP violation does not appear in the Yukawa couplings.
In a way, this is analogous to the case of a complex 
vacuum expectation value of the dilaton when CP gets broken via
the QCD $\theta$--term rather than the Yukawa couplings.

Another consequence of the above result is that
there is an additional contribution to the trilinear soft
SUSY breaking  parameters $A_{\alpha \beta \gamma}$
coming from the Yukawa coupling dependence on the Wilson moduli $a_i$:
\begin{equation}
\Delta A_{\alpha \beta \gamma}= F^{a_{_i}} \partial_{a_{_i}} 
\ln Y_{\alpha \beta \gamma}~\sim ~ {\cal O}(F^{a_{_i}} \bar a_{_i})~, 
\end{equation}
where $F^{a_{_i}}$ denotes the $F$-components of 
$a_i$ and 
we have taken into account the dominance of  the first term in the sum (\ref{result}).
Since one expects $a_i \sim$ ${\cal O}$(1) \cite{Bailin:1998yt}, 
generally this contribution is significant and should be included
in phenomenological analyses (see also \cite{Kim:1996ad}).
In particular, due to its flavor--dependence, this additional term is
relevant to the ``string CP problem'' \cite{Abel:2001cv}.

To conclude, we have studied the effect of the continuous Wilson lines
on the twisted Yukawa couplings.
Unlike in the open string case, the presence of the Wilson lines
affects the magnitude and hierarchy of the Yukawa couplings
instead of producing  phase factors.
These results have implications for the studies of CP violation
and supersymmetry breaking in heterotic string models. 

{\bf Acknowledgements.} We thank J. Erler and D. Jungnickel for
communications, and D. Bailin and T. Higaki for discussions.  
T.K. is supported in part by the Grants-in-Aid for 
Scientific Research No.14540256 from the Japan Society 
for the Promotion of Science.
O.L. is supported by PPARC.


\begin{thebibliography}{99}

\bibitem{Gross:1984dd}
D.~J.~Gross, J.~A.~Harvey, E.~J.~Martinec and R.~Rohm,
%``The Heterotic String,''
Phys.\ Rev.\ Lett.\  {\bf 54}, 502 (1985).

\bibitem{Dixon:1986qv}
L.~J.~Dixon, D.~Friedan, E.~J.~Martinec and S.~H.~Shenker,
%``The Conformal Field Theory Of Orbifolds,''
Nucl.\ Phys.\ B {\bf 282}, 13 (1987).

\bibitem{Hamidi:1986vh}
S.~Hamidi and C.~Vafa,
%``Interactions On Orbifolds,''
Nucl.\ Phys.\ B {\bf 279}, 465 (1987).

\bibitem{Burwick:1990tu}
T.~T.~Burwick, R.~K.~Kaiser and H.~F.~Muller,
%``General Yukawa Couplings Of Strings On Z(N) Orbifolds,''
Nucl.\ Phys.\ B {\bf 355}, 689 (1991).

\bibitem{erler}
J.~Erler, D.~Jungnickel, M.~Spalinski and S.~Stieberger,
%``Higher twisted sector couplings of Z(N) orbifolds,''
Nucl.\ Phys.\ B {\bf 397}, 379 (1993).

\bibitem{Ibanez:1986ka}
L.~E.~Ibanez,
%``Hierarchy Of Quark - Lepton Masses In Orbifold Superstring Compactification,''
Phys.\ Lett.\ B {\bf 181}, 269 (1986);
J.~A.~Casas and C.~Munoz,
%``Fermion Masses And Mixing Angles: A Test For String Vacua,''
Nucl.\ Phys.\ B {\bf 332}, 189 (1990)
[Erratum-ibid.\ B {\bf 340}, 280 (1990)];
J.~A.~Casas, F.~Gomez and C.~Munoz,
%``Complete structure of Z(n) Yukawa couplings,''
Int.\ J.\ Mod.\ Phys.\ A {\bf 8}, 455 (1993);
Phys.\ Lett.\ B {\bf 292}, 42 (1992);
A.~E.~Faraggi and E.~Halyo,
%``Cabibbo-Kobayashi-Maskawa mixing in superstring derived Standard - like Models,''
Nucl.\ Phys.\ B {\bf 416}, 63 (1994);
T.~Kobayashi,
%``Quark mass matrices in orbifold models,''
Phys.\ Lett.\ B {\bf 358}, 253 (1995);
T.~Kobayashi and Z.~z.~Xing,
%``A string-inspired ansatz for quark masses and mixing,''
Mod.\ Phys.\ Lett.\ A {\bf 12}, 561 (1997);
%``Quark mass matrices in superstring models,''
Int.\ J.\ Mod.\ Phys.\ A {\bf 13}, 2201 (1998);
J.~Giedt,
%``The KM phase in semi-realistic heterotic orbifold models,''
Nucl.\ Phys.\ B {\bf 595}, 3 (2001)
[Erratum-ibid.\ B {\bf 632}, 397 (2002)];
S.~A.~Abel and C.~Munoz,
%``Quark and lepton masses and mixing angles from superstring  constructions,''
arXiv:hep-ph/0212258.



\bibitem{Bailin:1998xx}
See, e.g., D.~Bailin, G.~V.~Kraniotis and A.~Love,
%``CP-violating phases in the CKM matrix in orbifold compactifications,''
Phys.\ Lett.\ B {\bf 435}, 323 (1998).

\bibitem{Lebedev:2001qg}
O.~Lebedev,
%``The CKM phase in heterotic orbifold models,''
Phys.\ Lett.\ B {\bf 521}, 71 (2001).


\bibitem{Aldazabal:2000sa}
G.~Aldazabal, L.~E.~Ibanez, F.~Quevedo and A.~M.~Uranga,
%``D-branes at singularities: A bottom-up approach to the string  embedding of the standard model,''
JHEP {\bf 0008}, 002 (2000);
S.~A.~Abel and A.~W.~Owen,
%``CP violation and CKM predictions from discrete torsion,''
Nucl.\ Phys.\ B {\bf 651}, 191 (2003).


\bibitem{Narain:am}
K.~S.~Narain, M.~H.~Sarmadi and E.~Witten,
%``A Note On Toroidal Compactification Of Heterotic String Theory,''
Nucl.\ Phys.\ B {\bf 279}, 369 (1987).

\bibitem{Ibanez:1986tp}
L.~E.~Ibanez, H.~P.~Nilles and F.~Quevedo,
%``Orbifolds And Wilson Lines,''
Phys.\ Lett.\ B {\bf 187}, 25 (1987).


\bibitem{Ibanez:1987xa}
L.~E.~Ibanez, H.~P.~Nilles and F.~Quevedo,
%``Reducing The Rank Of The Gauge Group In Oribifold Compactifications Of The Heterotic String,''
Phys.\ Lett.\ B {\bf 192}, 332 (1987) and refs. therein.


\bibitem{Dine:1992ya}
M.~Dine, R.~G.~Leigh and D.~A.~MacIntire,
%``Of CP and other gauge symmetries in string theory,''
Phys.\ Rev.\ Lett.\  {\bf 69}, 2030 (1992);
K.~w.~Choi, D.~B.~Kaplan and A.~E.~Nelson,
%``Is CP a gauge symmetry?,''
Nucl.\ Phys.\ B {\bf 391}, 515 (1993).


\bibitem{Kobayashi:1994ks}
T.~Kobayashi and C.~S.~Lim,
%``CP in orbifold models,''
Phys.\ Lett.\ B {\bf 343}, 122 (1995).


\bibitem{Cremades:2002va}
D.~Cremades, L.~E.~Ibanez and F.~Marchesano,
%``Towards a theory of quark masses, mixings and CP-violation,''
arXiv:hep-ph/0212064;
hep-th/0302105. 

\bibitem{Bailin:nk}
D.~Bailin and A.~Love,
%``Orbifold Compactifications Of String Theory,''
Phys.\ Rept.\  {\bf 315}, 285 (1999).

\bibitem{Kobayashi:1990mc}
T.~Kobayashi and N.~Ohtsubo,
%``Yukawa Coupling Condition Of Z(N) Orbifold Models,''
Phys.\ Lett.\ B {\bf 245}, 441 (1990).


\bibitem{Kobayashi:1991rp}
T.~Kobayashi and N.~Ohtsubo,
%``Geometrical aspects of Z(N) orbifold phenomenology,''
Int.\ J.\ Mod.\ Phys.\ A {\bf 9}, 87 (1994).



\bibitem{Kawai:1985xq}
H.~Kawai, D.~C.~Lewellen and S.~H.~Tye,
%``A Relation Between Tree Amplitudes Of Closed And Open Strings,''
Nucl.\ Phys.\ B {\bf 269}, 1 (1986).


\bibitem{Mohaupt:1993fb}
T.~Mohaupt,
%``Orbifold compactifications with continuous Wilson lines,''
Int.\ J.\ Mod.\ Phys.\ A {\bf 9}, 4637 (1994).

\bibitem{Cvetic:1995dt}
M.~Cvetic, B.~A.~Ovrut and W.~A.~Sabra,
%``Kahler potentials for orbifolds with continuous Wilson lines and the symmetries of the string action,''
Phys.\ Lett.\ B {\bf 351}, 173 (1995).

\bibitem{Dent:2001cc}
T.~Dent,
%``CP violation and modular symmetries,''
Phys.\ Rev.\ D {\bf 64}, 056005 (2001).


\bibitem{Khalil:2001dr}
S.~Khalil, O.~Lebedev and S.~Morris,
%``CP violation and dilaton stabilization in heterotic string models,''
Phys.\ Rev.\ D {\bf 65}, 115014 (2002);
O.~Lebedev and S.~Morris,
%``Towards a realistic picture of CP violation in heterotic string models,''
JHEP {\bf 0208}, 007 (2002).

\bibitem{Bailin:1998yt}
D.~Bailin, G.~V.~Kraniotis and A.~Love,
%``The effect of Wilson line moduli on CP-violation by soft supersymmetry  breaking terms,''
Phys.\ Lett.\ B {\bf 432}, 343 (1998).

\bibitem{Kim:1996ad}
H.~B.~Kim and C.~Munoz,
%``Orbifolds with continuous Wilson lines and soft terms,''
Mod.\ Phys.\ Lett.\ A {\bf 12}, 315 (1997).

\bibitem{Abel:2001cv}
S.~Abel, S.~Khalil and O.~Lebedev,
%``The string CP problem,''
Phys.\ Rev.\ Lett.\  {\bf 89}, 121601 (2002).


\end{thebibliography}
\end{document}